\documentclass[11 pt,final,journal,letterpaper,oneside,onecolumn]{IEEEtran}
\usepackage{multicol}   
\usepackage{multirow}
\usepackage[pdftex]{graphicx}
\usepackage[lined,boxruled]{algorithm2e}
\usepackage{algorithmic}
\usepackage[small]{caption}
\usepackage{float}
\usepackage{xspace}
\usepackage{fancyhdr}
\usepackage{amssymb,amsmath}
\usepackage{subfig}
\usepackage{textcomp} 
\usepackage{epstopdf}
\usepackage{bbding}
\usepackage{color}
\begin{document}

\title{\LARGE Propagation Channels for mmWave Vehicular Communications: State-of-the-art and Future Research Directions}

\author{Furqan Jameel, Shurjeel Wyne, Syed Junaid Nawaz, and Zheng Chang}


\maketitle

\begin{abstract}
Vehicular communications essentially support automotive applications for safety and infotainment. For this reason, industry leaders envision an enhanced role of vehicular communications in the fifth generation of mobile communications technology. Over the years, the number of vehicle-mounted sensors has increased steadily, which potentially leads to more volume of critical data communications in a short time. Also, emerging applications such as remote/autonomous driving and infotainment such as high-definition movie streaming require data-rates on the order of multiple Gbit/s. Such high data-rates require a large system bandwidth, but very limited bandwidth is available in the sub-6 GHz cellular bands. This has sparked research interest in the millimeter wave (mmWave) band (10 GHz-300 GHz), where a large bandwidth is available to support the high data-rate and low-latency communications envisioned for emerging vehicular applications. However, leveraging mmWave communications requires a thorough understanding of the relevant vehicular propagation channels, which are significantly different from those investigated below 6 GHz. Despite their significance, very few investigations of mmWave vehicular channels are reported in the literature. This work highlights the key attributes of mmWave vehicular communication channels and surveys the recent literature on channel characterization efforts in order to provide a gap analysis and propose possible directions for future research.
\end{abstract}


\IEEEpeerreviewmaketitle

\section{Introduction}
Vehicular communication networks have been extensively investigated to realize the concept of intelligent transportation systems (ITS). Such systems are motivated by the need to enhance safety through collision avoidance, provide driver-assistance through cruise-control and parking assistance, and alleviate traffic congestion \cite{shafi20175g}. For these reasons, vehicular communications constitute an important use-case of the fifth generation of mobile communications networks and many applications requiring high data-rate and low-latency transmission are planned \cite{shafi20175g}. Traditionally, vehicular communications have been supported by the dedicated short-range communication (DSRC) standard, which can ideally support up to 27 Mbit/s data-rates within a distance of 1 km. However, with the introduction of data-intensive automotive sensors such as laser imaging detection and ranging (LIDAR), visual/infrared cameras, and emerging infotainment applications the vehicular communications will be required to support Gbit/s data-rates. Such  high rates require a large system bandwidth but very limited bandwidth is available in the cellular bands below 6 GHz. These facts have motivated massive research interest in the millimeter wave (mmWave) frequency band (10 GHz-300 GHz) that has available large bandwidth to support high data-rate and low-latency transmissions. Table \ref{tab_2} shows some advantages of mmWave communications over the DSRC solutions like ITS-G5/DSRC and IEEE 802.11p/DSRC \cite{tassi2017modeling}.

\begin{figure*}[!htp]
\centering
\includegraphics[trim={0 6cm 0 2cm},clip,scale=.8]{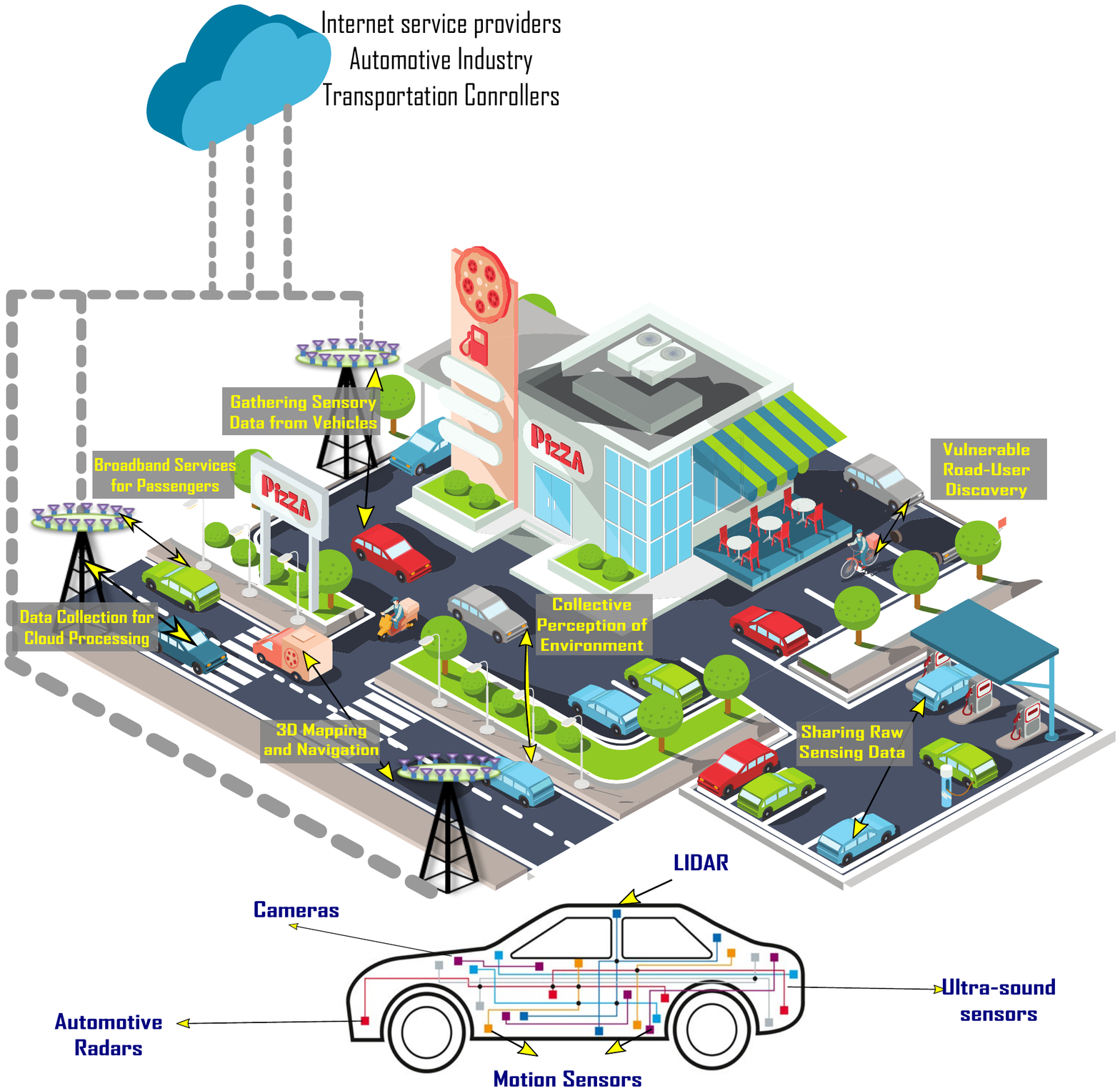}
\caption{Illustration of various applications of mmWave vehicular communications.}
\label{fig.1}
\end{figure*}

The mmWave band has been adopted for consumer electronics, e.g., in the 60 GHz unlicensed band the IEEE 802.11ad standard for wireless local area networks supports up to 7 Gbit/s data-rates. For the automotive industry the use of mmWave has been limited to automotive radars only. However, mmWave vehicular communications can potentially enable new applications of road safety and infotainment. For instance, one vehicle can share road information with neighboring vehicles in real-time to reveal sharp curves and objects through low-latency vehicle-to-vehicle (V2V) links. The vehicle-to-infrastructure (V2I) links can be used to gather information from the vehicles and forward it to some central traffic controller for congestion-free traffic regulation. The infrastructure can assist vehicle navigation by streaming high definition (HD) maps and live images of the streets through high data-rate low-latency mmWave links. The mmWave V2I links can also provide high speed internet to passing vehicles, which can then use intra-vehicle mmWave communications to broadcast the received data to its passengers. Some use scenarios are depicted in Fig. \ref{fig.1} and for further details see \cite{tassi2017modeling} and references therein.%
\begin{table}
\centering
\caption{Access technologies for vehicular networks \cite{tassi2017modeling}.}
\label{tab_2}
\begin{tabular}{|p{1.5cm}|p{2cm}|p{2cm}|p{1.5cm}|}
\hline
\textbf{Parameters} & \textbf{IEEE 802.11p/DSRC, ITS-G5/DSRC} & \textbf{LTE-A} & \textbf{MmWave systems} \\ \hline
Frequency Band& 5.85-5.925 GHz & 450 MHz - 4.99 GHz & 28, 38, 60 GHz \\ \hline
Channel Bandwidth& 10 MHz & Up to 100 MHz & 100 MHz - 2.16 GHz \\ \hline
Bit Rate& 3-27 Mbps & Up to 1 Gbps & Up to 7 Gbps \\ \hline
Latency& $\leq$ 10 ms & 100-200 ms & $\leq$ 10ms \\ \hline
Mobility Support& $\leq$ 130 km/h & $\leq$ 350 km/h & $\leq$ 100 km/h \\ \hline
\end{tabular}
\end{table}%

The mmWave propagation channels are significantly different from microwave channels because of the smaller signal wavelengths. The free-space pathloss varies inversely with the squared wavelength, which implies that mmWave signals experience more attenuation and therefore a reduced transmission range. Furthermore, objects like the human body and vehicles become large in comparison with the wavelength that leads to increased scattering. Also, mmWave signals suffer a high penetration loss when passing through common materials, which leads to frequent signal blockage and antenna placement on the vehicles becomes a design challenge due to this frequent blockage. Finally, the rapid temporal channel fluctuations caused by the vehicle mobility poses design challenges. However, despite the recent research interest in mmWave vehicular networks, only a few authors have focused on the underlying propagation channel \cite{shafi20175g}. This shortcoming is addressed in this survey, which has its main contributions listed as follows:%
\begin{itemize}
\item Identification of the key attributes of mmWave vehicular communication channels
\item Discussion of the interplay between mmWave channel characteristics and vehicular network design
\item Survey of recent mmWave vehicular channel investigations to provide a gap analysis and some recommendations for future measurements and analysis
\end{itemize}

The remainder of this paper is organized as follows. In Section II, the key attributes of mmWave vehicular channels are described. In Section III, we discuss the impact of channel characterization on mmWave vehicular network design. Section IV provides a survey of the state-of-the-art in mmWave vehicular channel characterization. Section V gives a taxonomy of the mmWave vehicular channels and provides some recommendations for future measurement and analysis. Finally the paper is concluded in Section VI.
\section{Key Characteristics of mmWave Channels: A Vehicular Communications Perspective}
\begin{figure*}[t]
\centering
\includegraphics[trim={0 8cm 0 0cm},clip,scale=.7]{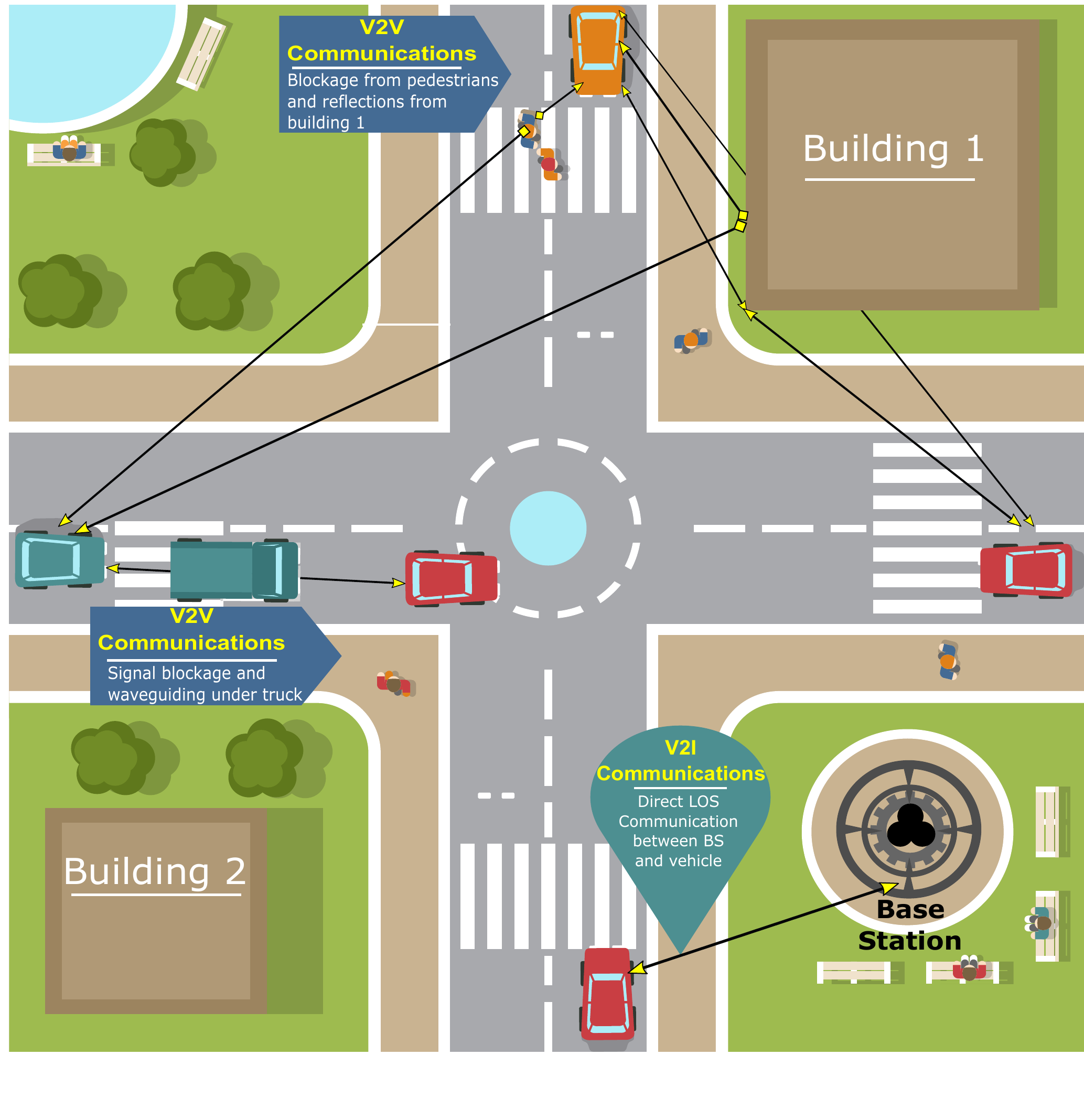}
\caption{Typical propagation effects and link types in mmWave vehicular communications.}
\label{fig.4}
\end{figure*}

As shown in Fig. \ref{fig.4}, mmWave vehicular links may experience a variety of channel conditions. For instance, a V2V signal may get blocked by a truck, but the gap between the truck and road surface may act as a waveguide. In addition, the V2V/V2I signals may be obstructed by adjacent buildings. Due to these reasons, line-of-sight (LOS) communications is difficult to maintain for the communicating vehicles. An appropriate channel model must account for the key attributes of mmWave vehicular channels, which are elaborated below. 
\subsection{Pathloss and blockage}
Pathloss refers to the distance-dependent attenuation of the received power. The Friis free-space equation reveals that the pathloss increases by 29 dB when the carrier frequency increases from 2 GHz to 60 GHz, using omnidirectional antennas. The mmWave signals also suffer high material penetration loss and poor diffraction capability; the inability of mmWaves to penetrate through urban canyon buildings has motivated the use of Manhattan distance-based pathloss models, discussed in Sec IV-A. Also, for vehicles moving on the innermost fast-lanes of a highway, the LOS to the serving base station (BS) at roadside may get blocked frequently by larger vehicles such as trucks moving on the outer slow-lanes, which needs to be modeled accurately \cite{tassi2017modeling}. Finally, for applications like mmWave radars for collision-avoidance in LOS scenarios, the grazing ground reflections carry significant power so that 2-ray pathloss modeling can be considered. Though in general, the application of the 2-ray model to vehicular communications is not straightforward due to frequent LOS blockage \cite{Anjinappa2018rayTracing}.
\subsection{Coherence time}
The propagation channel's coherence time is a statistical measure of the channel's time-rate of variability. It determines how rapidly the channel state information requires updates for system operation, e.g., re-alignment of antenna beams. The coherence time is inversely proportional to the channel's Doppler spread, which in turn increases linearly with the carrier frequency. Based on Clarke's model for the Doppler spectrum, one may assert that mmWave vehicular channels suffer from small values of coherence time. However, the Clarke's model assumes multipath components (MPCs) arriving uniformly in azimuth. But mmWave systems typically employ beamforming with very narrow beamwidths to increase the received signal strength and this directional reception can also reduce the Doppler spread, which has the desirable effect of increasing the coherence time \cite{7742901}.
\subsection{Clustering}
A cluster is generally defined as a set of MPCs having similar delays, angle-of-arrival (AoA), angle-of-departure (AoD), and Doppler spread \cite{gustafson2014mm}. Modeling the propagation channel in terms of MPC cluster parameters is more convenient than modeling the individual MPC parameters. Unlike existing cluster models such as the Saleh-Valenzula model and the WINNER model, the multipath clustering models for mmWave vehicular networks are under-explored. Owing to the channel sparsity of mmWave links, the models of the angular statistics at lower frequencies cannot be directly applied to mmWave links. Therefore, an accurate determination of MPC parameters and their cluster-based modeling requires special attention for mmWave vehicular channels.
\subsection{Spatial stationarity}
The assumptions of spatial stationarity and wide-sense stationary uncorrelated scattering are generally used to determine the MPC parameters like AoA/AoD, delay, and power. However, these assumptions may be violated for mmWave vehicular communications due to the availability of large bandwidth (up to 2 GHz) and rapid mobility of vehicles. The node mobility particularly introduces the \emph{spatial consistency} requirement for accurate channel modeling, i.e., the need to correctly model the joint channel parameters such as the pathloss at multiple locations along the mobile's trajectory \cite{8032491}. 
\section{Impact of Channel Characterization on mmWave Vehicular Network Design}
This section highlights the interplay between mmWave channel attributes and reliable vehicular network design.
\subsection{Infrastructure cost of vehicular networks}
Conventional models for the infrastructure cost are based on the coverage-to-capacity relationship. The cost to deploy mmWave BSs everywhere may not be feasible because mmWave links have relatively short communication range. Such links will most likely be deployed in urban environments or on bridges across highways, whereby the high vehicle speeds can be compensated by the relatively straight vehicular trajectories. An effective channel characterization can help in network design by determining appropriate BS placements to increase their coverage range and the number of simultaneously served vehicles.
\subsection{Antenna location}
Vehicles such as personal cars and public transport have different dimensions and mobility dynamics. The vehicle's antenna structure and its placement on the vehicle-body heavily impact the channel conditions on the communication link. While the mmWave band provides large array gains by allowing electrically large arrays in small physical dimensions, antenna placement is a design challenge due to the increased likelihood of mmWave signal blockage from the mounting vehicle's own body or from neighboring vehicles. Accurate mmWave vehicular channel characterization, considering vehicle-type, can help determine antenna-array design parameters like its form factor, element numbers, and placement on the vehicle-body. However, automotive companies may have different preferences for antenna-placement based on their vehicle aesthetics. Therefore, standardization efforts for antenna structures as well as signal processing techniques and waveform-related parameters require further work \cite{shafi20175g}.
\subsection{Multi-hop communications and routing protocols}
For vehicular networks, multi-hop communications and the underlying routing protocols play a key role in reliable message dissemination. In view of the increased pathloss and blockage probability of mmWave signals relative to microwave signals, the neighbor-selection procedure and the total hops have a greater impact on the performance of multi-hop mmWave vehicular networks. As the neighboring vehicle-density and the radio environment can rapidly vary in vehicular networks, the characterization of mmWave multi-hop channels and their link correlations is necessary for optimum design of routing protocols for relaying-based mmWave vehicular networks.
\subsection{Beam-alignment}
The mmWave systems typically employ antenna beamforming at transmitter and receiver to overcome the increased pathloss. However, such directional links require a precise beam-alignment. In case of multiple switched beams, the appropriate transmit and receive beams need to be sequentially searched through a codebook, which introduces beam-alignment delay. In particular, the joint effect of the vehicle's changing travel-directions and narrow mmWave beams can cause a rapid loss of alignment resulting in link outage. While compressive channel estimation techniques have been proposed to exploit the mmWave channel sparsity, their overhead is still too high for mobility scenarios. As mmWave systems will typically coexist with sub-6 GHz networks, the sensor information from different bands can be utilized for the application of beam-alignment in mmWave vehicular communications \cite{Prelcic2017outBandComm}.
\subsection{Coexistence of mmWave and microwave links}
In the near future, mmWave systems are expected to be used in conjunction with sub-6 GHz networks and the joint scheduling of mmWave and microwave links will be a design issue. For a fixed number of users, the resource allocation schemes would have to address how the resource allocation over mmWave links will affect the microwave links and vice versa \cite{7929424}. In addition, the quality-of-service requirements will dictate whether the services will be provided either through one of the microwave/mmWave resources or jointly.
\section{State-of-the-Art of Channel Modeling for MmWave Vehicular Networks}
This section surveys some recent experimental and analytical channel models for mmWave vehicular networks, which are classified into mainly three categories, viz: V2I channels, Inter-vehicle channels, and Intra-vehicle channels.
\subsection{V2I channels}
The mmWave V2I communications can potentially realize high data-rate low-latency links for different mobile devices within the network. The mmWave urban vehicular networks, excluding their mobility aspect and lower height of roadside BSs, bear a close resemblance to the well-explored mmWave urban microcellular (UMi) networks \cite{8032491}. Both communication systems are characterized by a dense network of streets and high-rise structures wherein the mmWave signals are attenuated while propagating through street-canyons and diffracting along street corners - these signals do not penetrate the neighboring buildings, which justifies the use of the link's Manhattan distance rather than its Euclidean distance for analyzing the pathloss \cite{wang2018mmwave}.

In \cite{8032491} the authors analyzed the suitability of traditional pathloss modeling methods for 28 GHz UMi environment by using a calibrated ray-tracing data-set for New York City (NYC).\footnote{The traditional modeling of power-law pathloss summed with log-normal variations considers the pathloss data across multiple streets as one ensemble to determine model parameters.} Their analysis verified that the mmWave pathloss is non-stationary in that it does not depend solely on the absolute link distance and that by using traditional pathloss models, the shadowing-variance is over-estimated. They proposed a spatially consistent street-by-street pathloss model for the 28 GHz UMi environment, which treats the pathloss parameters as random variables modeled on a street-by-street basis while taking into account the street-orientation. However, these results may also be corroborated with measurement-based analysis of street-canyon environments in other major cities.

The authors in \cite{wang2018mmwave} proposed a Manhattan distance-based pathloss model for mmWave urban V2I communications. Their model uses the LOS blockage probability as a key parameter for deriving the system's coverage probability. Their analysis showed that the Manhattan distance-based pathloss models can have quite different coverage probability and ergodic capacity compared with those evaluated from Euclidean distance-based pathloss models. Their results also revealed that the LOS interference from BSs co-located with the serving BS is the dominant factor in determining the coverage performance and that the non-LOS BSs mostly have an insignificant contribution to the interference and the association probability. Finally, the authors also showed that the coverage probability levels off asymptotically in the number of BSs, whereas an increase in the street-intensity for dense BS deployments reduces the coverage probability but for smaller BS intensities an increase in the street-intensity also increases the coverage probability. Therefore, only a limited number of BSs need to be deployed in a dense urban street environment.

The efficient design of channel-estimation algorithms requires an accurate characterization of the channel's coherence-time, i.e., the time over which the channel can be considered invariant. Though mmWave beamforming can compensate the pathloss through narrow beamwidths, these pencil-like beams also increase the beam-pointing errors under mobility. In \cite{7742901} the authors derived closed-form expressions to describe the channel coherence time as a function of the beamwidth in the presence of beam-alignment errors and Doppler effect. Their analysis showed the existence of a non-trivial optimal beamwidth to maximize the coherence time. A narrower beamwidth reduces the coherence time due to increased beam-pointing errors, whereas a broader-than-optimal beamwidth increases the Doppler frequency shift, which also reduces the coherence time. This finding is of high significance in the design of dynamic sub-carrier spacing in multi-carrier communication systems.  
 
Beam-alignment also requires an accurate determination of vehicle location in order to identify the beam-pointing directions. The authors in \cite{va2017inverse} developed a position-aided channel model for a realistic performance evaluation of beam-alignment techniques. Their simulation setup considered a two-lane urban street environment in which the roads and buildings were made of asphalt and concrete, respectively. A geometric spatial channel model was used to improve the modeling precision of the MPC statistics that were drawn from a ray-tracing simulator. A long-term multipath fingerprinting database (which is the inverse of fingerprinting localization) was constructed for dense traffic conditions. It was demonstrated that the spatial side-information such as the receiver location can be leveraged to predict the optimal beam-pair index based on the database of previous beam training results. 
\subsection{Inter-vehicle channels}
Characterization of inter-vehicle channels is important to ensure reliable communications between the vehicles. For mmWave V2V links, the channel characteristics such as the total number of MPC clusters and their intra-cluster scattering determines the channel's Doppler statistics, which in turn describe the propagation channel's temporal variability \cite{he2017geometrical}. The signal scattering is caused by structural discontinuities that can occur on the otherwise smooth reflecting surface of neighboring vehicles and roadside objects. Also, for V2V communications the channel's root-mean-squared (RMS) delay-spread typically reduces for increasing signal frequencies due to the higher attenuation and scattering at the smaller wavelengths \cite{rappaport2015wideband}.

In \cite{he2017geometrical}, the authors proposed a multiple-input multiple-output (MIMO) two-ring geometric channel model for mmWave V2V communications. They characterized the Doppler spectrum, power delay profile (PDP), and time-frequency correlation function for the considered scenario. The channel model of \cite{he2017geometrical} is generalized in that it can adapt to different mmWave radio propagation scenarios by tuning the position and number of effective scattering clusters and the degree of their scattering effect. Their analysis reveals that a small number of scattering clusters leads to a higher channel correlation. Also, the use of directional antennas, which reduces the Doppler spread, increases the channel correlation relative to that observed by employing omnidirectional antennas. The authors propose to deploy directional antennas at both the back- and front-side of vehicles to ensure reliable connections for safety-critical applications.
 
In \cite{Petrov2018_V2V_Intrf}, the authors used 79 GHz V2V measurements and stochastic geometry to analyze the interference in 3-lane highway and urban scenarios. Specifically, they considered the interference caused by vehicles on the side-lanes to a vehicle traveling on the center-lane. All vehicles were considered to have front- and rear-mounted directional antennas for communicating with neighboring vehicles. The authors observed that antenna beamwidths less than 20 degrees captured minimal interference from the side-lanes such that interference management schemes were not required. However, the larger angular coverage achieved by beamwidths greater than 40 degrees also captured significant interference from the side-lanes, which necessitated the use of interference mitigation schemes for reliable communications.

In \cite{rappaport2015wideband}, the authors carried out extensive mmWave vehicle-to-everything (V2X) channel measurements in the cities of NYC and Austin for the 28, 38, 60, and 73 GHz bands. Their measurements revealed larger pathloss exponents, on average, in NYC than those observed in Austin, due to the higher density of urban structures in NYC. Furthermore, the pathloss exponent for a given environment was shown to increase with increasing center-frequency due to the increased scattering at smaller wavelengths. For Austin, a typical less urban city, the propagation channel's RMS delay-spread was observed to decrease linearly with increasing link distance. However, for a dense urban environment like NYC, the RMS delay-spread was observed to be significantly larger due to the presence of highly reflective urban structures. The mmWave links measured in Austin were observed to have a higher coverage probability than that of similar links measured in NYC due to an increased signal-blockage probability by the dense urban structures in NYC. However, mmWave communications were deemed possible at ranges below 200 m in a majority of the measured links. The authors also proposed pathloss models, based on their measurements, for several mmWave bands and a unified statistical channel impulse-response model for these mmWave bands.
\subsection{Intra-vehicle channels}
The position of communicating nodes in intra-vehicle propagation environments is usually considered to be relatively stationary. However, the impact of vehicle vibrations on the radio link requires a comprehensive investigation. These vibrations are caused by various physical factors such as engine vibrations (dependent on its temperature, structure, and revolution-rate), in-vehicle audio entertainment system, and road surface condition. For a typical 5 mm wavelength, these vibrations can cause a severe Doppler spread, which leads to rapid fluctuations of the received signal envelope and may cause performance degradation of the communication system. In \cite{blumenstein2016effects}, the effect of vibrations, caused by road conditions and vehicle velocity, on the Doppler spread of the intra-vehile channel is investigated. The authors recorded $3000$ temporal snapshots each of more than $30$ channel impulse responses (CIRs). Their investigations revealed that the mean Doppler spread can reach up to 38 Hz, while the number of resolvable multipath clusters was observed to be $4$. Though some very useful insights were provided in \cite{blumenstein2016effects}, there is still a need to conduct more studies to quantify the effect of vehicle vibrations on the Doppler spread of mmWave intra-vehicle channels for different engine-types.

In \cite{blumenstein2017vehicle}, a comprehensive measurement-based investigation was conducted to compare intra-vehicle communications at the microwave (3-11 GHz) and mmWave (55-65 GHz) bands. The microwave measurements were made with omnidirectional antennas, whereas the mmWave measurements were conducted with directive antennas pointing towards each other. The channel frequency responses were recorded with a vector network analyzer with a frequency resolution of 70 MHz. The pathloss was measured to be 36 dB and 52 dB for the 3-11 GHz and 55-65 GHz bands, respectively. However, the pathloss-variance for both bands was observed to be similar and around 20 dB$^2$. The RMS delay-spreads were measured to be 31 ns and 13 ns at 3 and 11 GHz, respectively, whereas RMS delay-spreads of 24 ns and 10 ns were observed at 55 GHz and 65 GHz, respectively. The delay-spread variance for the 3-11 GHz band was measured around 10 ns$^2$, which was considerably greater than the 5 ns$^2$ delay-spread variance observed for the 55-65 GHz band. The authors proposed a piece-wise linear PDP model for both the 3-11 and 55-65 GHz frequency bands. The model's validity was demonstrated by applying the Kolmogorov-Smirnov goodness-of-fit test. The 55-65 GHz band was observed to have a higher PDP collinearity than the 3-11 GHz band. The CIR realizations for a $10\times 10$ antenna grid were averaged to analyze the spatial stationarity and the number of resolvable MPC clusters in the measured intra-vehicle channels. The number of MPC clusters was observed to be 2 for the 3-11 GHz band and between 3-5 for the 55-65 GHz band, which indicates a more prominent clustering behavior for the latter.
\section{Taxonomy of MmWave Vehicular Channels and Future Recommendations}
\subsection{Taxonomy}
\begin{figure*}[t]
\centering
\includegraphics[trim={0 6cm 0 6cm},clip,scale=.8]{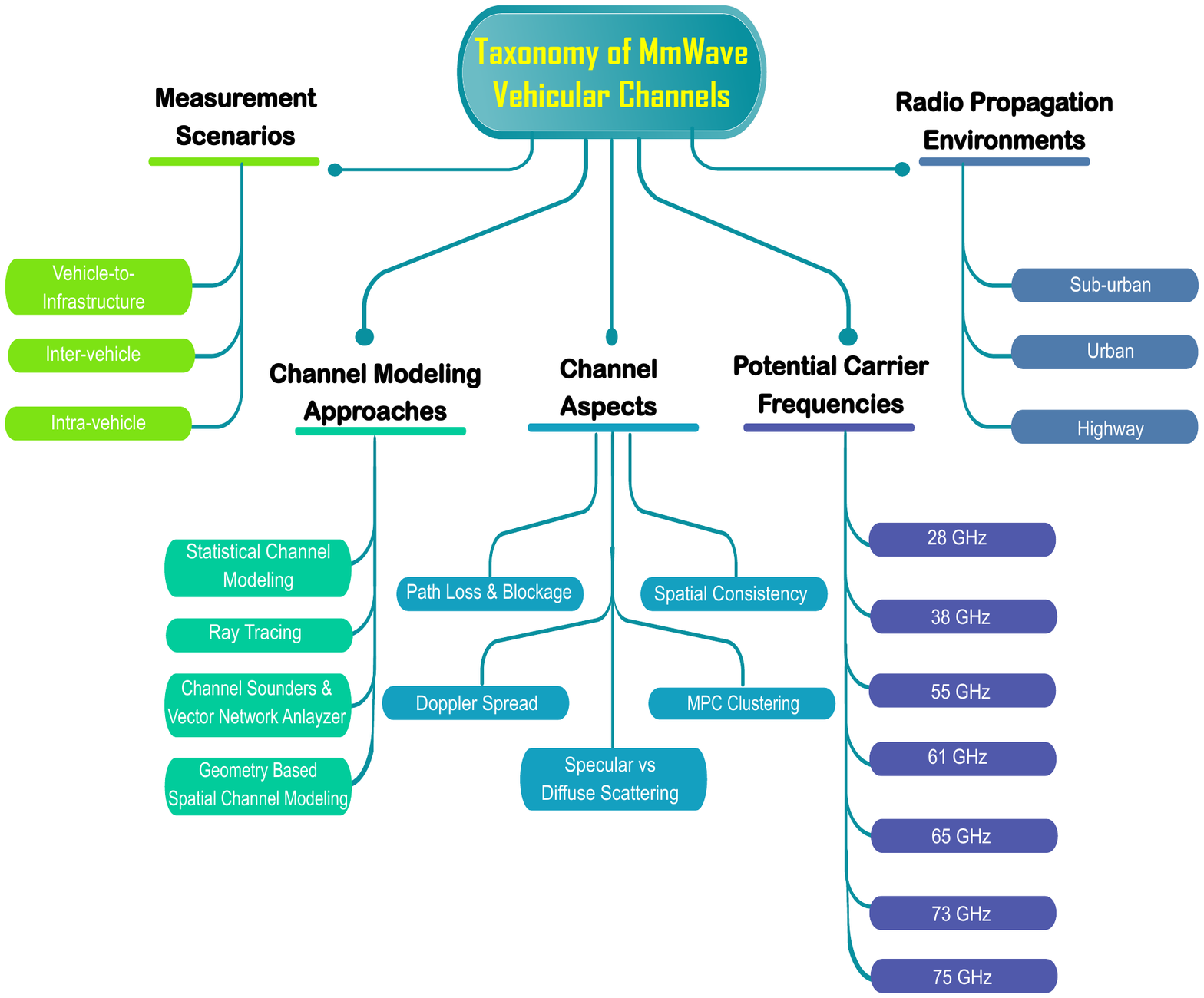}
\caption{Taxonomy of mmWave vehicular channels.}
\label{fig.2}
\end{figure*}

In light of the propagation-related investigations discussed above, a brief taxonomy of mmWave vehicular channels is illustrated in Fig. \ref{fig.2}. The notable approaches for channel modeling are ray-tracing based, measurement-based, statistical, and geometry-based stochastic channel modeling. Each of these approaches has its own advantages and limitations. 

The geometry-based modeling approach establishes a probabilistic relationship between the spatial positions of the transmitter, receiver, and scattering objects but lacks in accuracy for modeling a specific propagation environment. On the other hand, the ray-tracing and measurement-based approaches offer more precision in modeling a specific propagation scenario but are applicable only to that scenario for which the campaign/simulation is conducted. The heavy computational requirements of the ray-tracing approach limits its applicability for real-time channel predictions. Another limitation of ray-tracing, despite the availability of geometry files for some street canyons, is the difficulty to accurately model the physical aspects such as building material and mobility of objects etc. Therefore, ray-tracing may be used in an auxiliary role to channel measurements or it can be applied to the scenarios where long-term channel statistics are required. A few ray-tracing based channel models are reported in the literature for prediction of angular statistics and dominant clusters in mmWave vehicular communication environments. In the pure statistical modeling approach, the measured instances of the CIR or its frequency response are used as the statistical ensemble for extracting statistical parameters such as the RMS delay-spread. 

A few geometric channel models have been proposed in the literature for predicting the pathloss under various mmWave propagation conditions. However, more research efforts can be focused on modeling the channel profile jointly in the delay and angular domains. The channel measurement campaigns have been mostly conducted in urban or suburban environments, e.g., the comparative study between the channel statistics in the cities of NYC and Austin. For the intra-vehicle channel characterization, it may be difficult to perform accurate ray-tracing analysis due to the difficulty in correctly modeling the variety of materials used in car-interiors. Therefore, measurement campaigns can be more suitable for in-vehicle propagation channel characterization owing to the small area and relatively easy installation of transceivers. Further efforts may also be focused on characterizing mmWave propagation channels for roundabouts, bridges, tunnels, and multi-level highways. In these scenarios, the mmWave vehicular communications can be suitable for provisioning of cooperative awareness and localization services due to the availability of reasonable user density and geometric simplicity for modeling these structures.
\subsection{Future research directions}
Fig. \ref{fig.3} gives an overview of some future directions for channel modeling of mmWave vehicular links, which are further elaborated as follows: 
\begin{figure*}[t]
\centering
\includegraphics[trim={0 6cm 0 6cm},clip,scale=.8]{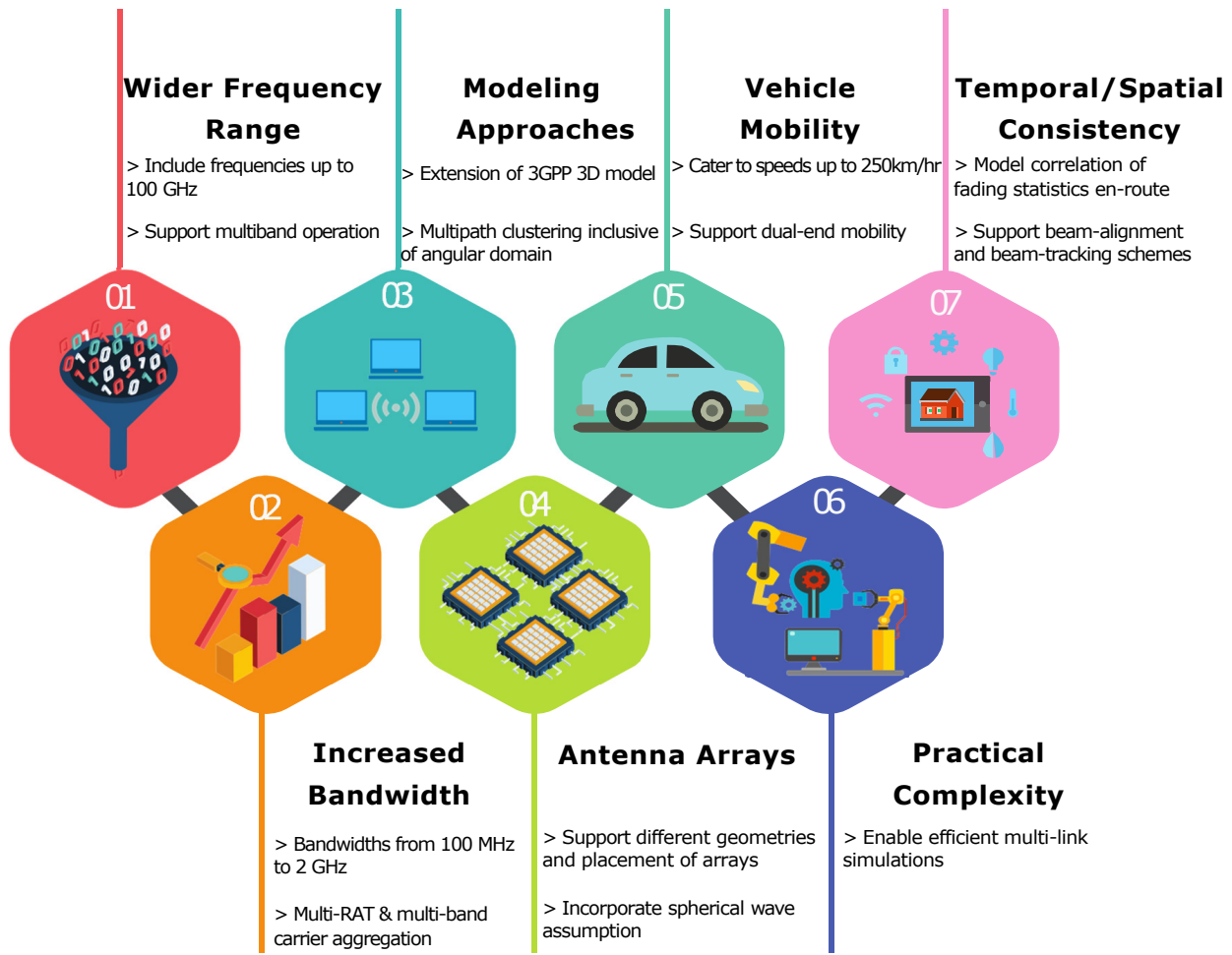}
\caption{Future considerations for mmWave vehicular channel modeling.}
\label{fig.3}
\end{figure*}
\begin{itemize}
\item The emerging applications of mmWave vehicular communications require channel models for a wide range of frequencies between 10 to 300 GHz. Moreover, to investigate the feasibility of multi-band vehicular communications, the joint propagation characteristics for high- and low-band carrier frequencies may be studied. Table \ref{tab_3} lists some environment-based considerations in planning measurement campaigns for mmWave vehicular links. 
\item The channel models for mmWave vehicular communications need to address the issue of channel bandwidth division. It is also anticipated that this increase in bandwidth may support carrier aggregation methods like multi-band and multi-radio access technology.
\item The existing 3GPP 3D channel modeling approaches of WINNERII, WINNER+, and 3GPP TR36.873 may be evolved to incorporate mmWave vehicular channels to cater to the requirements of future vehicular communication applications. The precise modeling of azimuth and elevation AoA and AoD need to be accounted for to achieve the required precision in beamforming. 
\item The design of accurate channel models for mmWave vehicular links may require a hybrid approach combining measurements, ray-tracing, and geometric modeling. In this context, the research community may benefit by generous policies for sharing the measurement/ray-tracing data-sets of mmWave vehicular channels. 
\item The mmWave channel models for vehicular communications must support large-scale antenna arrays, with different geometries such as linear, planar etc. It is important to mention that the manifold assumption may not hold for mmWave communications as the bandwidth is a significant fraction of the center-frequency.
\item For high-speed vehicles, the mmWave vehicular channel models need to support vehicle mobility with speeds up to 250 km/hr. Beam-tracking becomes challenging for inter-vehicle communications owing to the mobility of both transmitter and receiver. Thus, the future channel models need to consider the variations in travel-directions of both the transmitting and receiving vehicles for beam-tracking in mmWave vehicular networks.
\begin{table}
\centering
\caption{Key environment considerations for future measurement campaigns.}
\label{tab_3}
\begin{tabular}{|p{1.5cm}|p{2cm}|p{4cm}|}
\hline
\textbf{Measurement Environment} & \textbf{Characteristics}          & \textbf{Key Factors}\\ \hline
Mountainous/ Hilly               & Terrain, foliage, soil, rain      & Terrain height-variation and
slope, foliage density and type (trees/bushes), soil-moisture and ground-roughness, rain intensity.\\ \hline
Rural                            & Buildings, foliage                & Building height, type, and construction-material; foliage density and type; surface roughness.\\ \hline
Urban/ Suburban                  & Buildings, open area, foliage     & Building density, height, type, and construction-material; road size; numbers, locations, and heights of in-band and out-of-band BSs.\\ \hline
\end{tabular}
\end{table}
\item The future mmWave vehicular channel models should have manageable complexity to support multilink simulations without loss of accuracy. 
\item The proposed channel models for mmWave vehicular communications must support the rapid transitions between LOS and non-LOS conditions, which will be required for beam-tracking and radar-aided beam-alignment in V2I communications.
\end{itemize}
\section{Conclusion}
The mmWave band has emerged as a viable solution for high data-rate and low-latency vehicular communications. However, mmWave vehicular propagation channels are significantly different from those investigated at conventional cellular bands below 6 GHz. This work has identified key aspects of mmWave vehicular channels and highlighted the interplay between mmWave vehicular channel characteristics and the network design. A comprehensive review of the state-of-the-art of mmWave vehicular channel characterization has been presented to identify critical challenges and research opportunities that merit further investigation to improve the performance of future mmWave vehicular communication networks.
\section*{Acknowledgment}
This work is supported by the EU-funded project ATOM-690750, approved under call H2020-MSCA-RISE-2015.
\ifCLASSOPTIONcaptionsoff
  \newpage
\fi
\bibliographystyle{IEEEtran}
\bibliography{WCM-18-00174_Ref}

\begin{thebibliography}{10}
\providecommand{\url}[1]{#1}
\csname url@samestyle\endcsname
\providecommand{\newblock}{\relax}
\providecommand{\bibinfo}[2]{#2}
\providecommand{\BIBentrySTDinterwordspacing}{\spaceskip=0pt\relax}
\providecommand{\BIBentryALTinterwordstretchfactor}{4}
\providecommand{\BIBentryALTinterwordspacing}{\spaceskip=\fontdimen2\font plus
\BIBentryALTinterwordstretchfactor\fontdimen3\font minus
  \fontdimen4\font\relax}
\providecommand{\BIBforeignlanguage}[2]{{%
\expandafter\ifx\csname l@#1\endcsname\relax
\typeout{** WARNING: IEEEtran.bst: No hyphenation pattern has been}%
\typeout{** loaded for the language `#1'. Using the pattern for}%
\typeout{** the default language instead.}%
\else
\language=\csname l@#1\endcsname
\fi
#2}}
\providecommand{\BIBdecl}{\relax}
\BIBdecl

\bibitem{shafi20175g}
M.~Shafi, A.~F. Molisch, P.~J. Smith, T.~Haustein, P.~Zhu, P.~D. Silva,
  F.~Tufvesson, A.~Benjebbour, and G.~Wunder, ``{5G: A tutorial overview of
  standards, trials, challenges, deployment, and practice},'' \emph{IEEE
  Journal on Selected Areas in Communications}, vol.~35, no.~6, pp. 1201--1221,
  Jun. 2017.

\bibitem{tassi2017modeling}
A.~Tassi, M.~Egan, R.~J. Piechocki, and A.~Nix, ``{Modeling and design of
  millimeter-wave networks for highway vehicular communication},'' \emph{IEEE
  Transactions on Vehicular Technology}, vol.~66, no.~12, pp. 10\,676--10\,691,
  Dec. 2017.

\bibitem{Anjinappa2018rayTracing}
C.~K. Anjinappa and I.~Guvenc, ``Millimeter-wave {V2X} channels: propagation
  statistics, beamforming, and blockage,'' in \emph{Proc. {IEEE} Vehicular
  Technology Conference {(VTC2018-Fall)}}, Chicago, USA, Aug. 2018, pp. 1--6.

\bibitem{7742901}
V.~Va, J.~Choi, and R.~W. Heath, ``{The impact of beamwidth on temporal channel
  variation in vehicular channels and its implications},'' \emph{IEEE
  Transactions on Vehicular Technology}, vol.~66, no.~6, pp. 5014--5029, Jun.
  2017.

\bibitem{gustafson2014mm}
C.~Gustafsson, K.~Haneda, S.~Wyne, and F.~Tufvesson, ``On mm-wave multipath
  clustering and channel modeling,'' \emph{IEEE Transactions on Antennas and
  Propagation}, vol.~62, no.~3, pp. 1445--1455, Mar. 2014.

\bibitem{8032491}
A.~Karttunen, A.~F. Molisch, S.~Hur, J.~Park, and C.~J. Zhang, ``{Spatially
  consistent street-by-street path loss model for 28-GHz channels in micro cell
  urban environments},'' \emph{IEEE Transactions on Wireless Communications},
  vol.~16, no.~11, pp. 7538--7550, Nov. 2017.

\bibitem{Prelcic2017outBandComm}
N.~Gonzalez-Prelcic, A.~Ali, V.~Va, and R.~W. Heath, ``Millimeter-wave
  communication with out-of-band information,'' \emph{IEEE Communications
  Magazine}, vol.~55, no.~12, pp. 140--146, Dec. 2017.

\bibitem{7929424}
O.~Semiari, W.~Saad, and M.~Bennis, ``{Joint millimeter wave and microwave
  resources allocation in cellular networks with dual-mode base stations},''
  \emph{IEEE Transactions on Wireless Communications}, vol.~16, no.~7, pp.
  4802--4816, Jul. 2017.

\bibitem{wang2018mmwave}
Y.~Wang, K.~Venugopal, R.~W. Heath, and A.~F. Molisch, ``{MmWave
  vehicle-to-infrastructure communication: analysis of urban microcellular
  networks},'' \emph{IEEE Transactions on Vehicular Technology}, 2018.

\bibitem{va2017inverse}
V.~Va, J.~Choi, T.~Shimizu, G.~Bansal, and R.~W. Heath, ``Inverse multipath
  fingerprinting for millimeter wave {V2I} beam alignment,'' \emph{IEEE
  Transactions on Vehicular Technology}, vol.~67, no.~5, pp. 4042--4058, May
  2018.

\bibitem{he2017geometrical}
R.~He, B.~Ai, G.~L. St\"uber, G.~Wang, and Z.~Zhong, ``Geometrical-based
  modeling for millimeter-wave {MIMO} mobile-to-mobile channels,'' \emph{IEEE
  Transactions on Vehicular Technology}, vol.~67, no.~4, pp. 2848--2863, Apr.
  2018.

\bibitem{rappaport2015wideband}
T.~S. Rappaport, G.~R. MacCartney, M.~K. Samimi, and S.~Sun, ``{Wideband
  millimeter-wave propagation measurements and channel models for future
  wireless communication system design},'' \emph{IEEE Transactions on
  Communications}, vol.~63, no.~9, pp. 3029--3056, Sep. 2015.

\bibitem{Petrov2018_V2V_Intrf}
V.~Petrov, J.~Kokkoniemi, D.~Moltchanov, J.~Lehtom\"aki, M.~Juntti, and
  Y.~Koucheryavy, ``The impact of interference from the side lanes on
  {mmWave/THz band V2V} communication systems with directional antennas,''
  \emph{IEEE Transactions on Vehicular Technology}, vol.~67, no.~6, pp.
  5028--5041, Jun. 2018.

\bibitem{blumenstein2016effects}
J.~Blumenstein, J.~Vychodil, M.~Pospisil, T.~Mikulasek, and A.~Prokes,
  ``{Effects of vehicle vibrations on mm-wave channel: Doppler spread and
  correlative channel sounding},'' in \emph{Proc. IEEE Annual International
  Symposium on Personal, Indoor, and Mobile Radio Communications (PIMRC)},
  Valencia, Spain, Sep. 2016, pp. 1--5.

\bibitem{blumenstein2017vehicle}
J.~Blumenstein, A.~Prokes, A.~Chandra, T.~Mikulasek, R.~Marsalek, T.~Zemen, and
  C.~Mecklenbr\"auker, ``{In-vehicle channel measurement, characterization, and
  spatial consistency comparison of $\text{3}\hbox{--}\text{11 GHz}$ and
  $\text{55}\hbox{--}\text{65 GHz}$ frequency bands},'' \emph{IEEE Transactions
  on Vehicular Technology}, vol.~66, no.~5, pp. 3526--3537, May 2017.

\end{thebibliography}
\end{document}